\newif\iffmcad
\newif\ifnotfmcad
\fmcadtrue
\iffmcad\notfmcadfalse\else\notfmcadtrue\fi
\newif\ifarxiv
\newif\ifnotarxiv
\arxivtrue
\ifarxiv\notarxivfalse\else\notarxivtrue\fi
\iffmcad
  \documentclass[10pt,conference,letterpaper]{IEEEtran}

  \newtheorem{example}{Example}{\itshape}{\rmfamily}
  \newtheorem{theorem}{Theorem}{\itshape}{\rmfamily}
  \newtheorem{definition}{Definition}{\itshape}{\rmfamily}
\else
  \documentclass{llncs}
\fi

\usepackage{balance}
\usepackage{cite}
\usepackage[utf8]{inputenc}
\usepackage[T1]{fontenc}
\usepackage{amsmath}
\usepackage{amssymb}
\usepackage{bussproofs}
\usepackage{url}
\usepackage{while}
\usepackage{todonotes}

\iffmcad
  \renewcommand{\IEEEQED}{\IEEEQEDopen}
\fi

\usepackage{rotating}
\usepackage{multirow}

\usepackage{graphicx}
\usepackage{subcaption}
\usepackage{floatflt}
\usepackage{times}

\usepackage{xcolor}
\definecolor{mygreen}{RGB}{0,159,57}
\definecolor{myyellow}{RGB}{255,200,0}
\definecolor{myorange}{RGB}{255,140,0}
\colorlet{myred}{red!90}
\definecolor{myblue}{RGB}{0,50,255}

\newcommand{\tracelogic}{\mathcal{L}}

\newcommand{\tpsort}{\mathbb{L}}
\newcommand{\intsort}{\mathbb{I}}
\newcommand{\natsort}{\mathbb{N}}
\newcommand{\tracesort}{\mathbb{T}}

\newcommand{\Nat}{\mathbb{N}}
\newcommand{\Int}{\mathbb{I}}

\newcommand{\natsig}{S_{\Nat}}
\newcommand{\intsig}{S_{\Int}}
\newcommand{\tpsig}{S_{Tp}}
\newcommand{\lastsig}{S_n}
\newcommand{\pvsig}{S_V}
\newcommand{\tracesig}{S_{Tr}}

\newcommand{\pv}[1]{\texttt{#1}}
\newcommand{\pvi}[1]{$\mathtt{_{#1}}$}
\newcommand{\eql}{{\,\simeq\,}}
\newcommand{\neql}{\not\simeq}
\newcommand{\zero}{{\tt 0}}
\newcommand{\suc}{{\tt succ}}
\newcommand{\pred}{{\tt pred}}

\newcommand{\combTheory}{{\Nat\cup\Int}}

\usepackage{ stmaryrd }

\newcommand{\while}[1]{\lstinline[mathescape]!#1!}

\newcommand{\whilelang}{\mathcal{W}}

\newcommand{\ifThenElseStatement}{\text{if-then-else}}
\newcommand{\whileStatement}{\text{while}}

\newcommand{\rapid}{\textsc{Rapid}}
\newcommand{\vampire}{\textsc{Vampire}}
\newcommand{\avatar}{\textsc{Avatar}}
\newcommand{\smtlib}{\textsc{smt-lib}}

\usepackage{stackengine}
\stackMath

\newcommand{\weirdarrow}[1]{= \hspace{-0.5em}\gg_{#1}}

\newcommand{\whileOp}[3][]{\texttt{while}^{#1} \texttt{(} #2 \texttt{)do\{} #3 \texttt{\}}}
\newcommand{\iteOp}[3]{\texttt{if(} #1 \texttt{)then\{} #2 \texttt{\}else\{} #3 \texttt{\}}}
\newcommand{\skipOp}{\texttt{skip}}

\iffmcad
  \newcommand\qed{\vspace*{-1em}\hfill {\small $\Box$}}
\fi

\begin{document}

\title{
  Verifying Relational Properties using Trace Logic
  %and
%Superposition Proving%a Theorem Prover
}

\iffmcad
  \author{
    \IEEEauthorblockN{Gilles Barthe\IEEEauthorrefmark{1}\IEEEauthorrefmark{2},
      Renate Eilers\IEEEauthorrefmark{3},
      Pamina Georgiou\IEEEauthorrefmark{3},
      Bernhard Gleiss\IEEEauthorrefmark{3},  
      Laura Kov{\'a}cs\IEEEauthorrefmark{3}\IEEEauthorrefmark{4},  
      Matteo Maffei\IEEEauthorrefmark{3}
    }
    \IEEEauthorblockA{\IEEEauthorrefmark{1}Max Planck Institute for Security and Privacy, Germany}
    \IEEEauthorblockA{\IEEEauthorrefmark{2}IMDEA Software Institute, Spain}
    \IEEEauthorblockA{\IEEEauthorrefmark{3}TU Wien, Austria}
    \IEEEauthorblockA{\IEEEauthorrefmark{4}Chalmers University of Technology, Sweden}
  }
\else
  \author{
    Gilles Barthe\inst{1\and 2}
    \and
    Renate Eilers\inst{3}
    \and
    Pamina Georgiou\inst{3}
    \and
    Bernhard Gleiss\inst{3}
    \and
    Laura Kov{\'a}cs\inst{3\and 4}
    \and
    Matteo Maffei\inst{3}
  }

  \institute{
    Max Planck Institute for Security and Privacy, Germany
    \and
    IMDEA Software Institute, Spain
    \and 
    TU Wien, Austria
    \and
    Chalmers University of Technology, Sweden
  }
\fi

\maketitle        

\begin{abstract}
  We present a %novel
  logical framework for the verification of
  relational properties in imperative programs.
%
%Our work is motivated by relational properties which come from security applications and often require reasoning about formulas with quantifier-alternations.
%
  Our framework reduces verification of relational properties of
  imperative programs to a validity problem in trace logic, an
  expressive instance of first-order predicate logic. Trace logic
  draws its expressiveness from its syntax, which allows expressing
  properties over computation traces.
  Its axiomatization supports fine-grained reasoning about
  intermediate steps in 
  program execution, notably loop iterations.
  We present an algorithm to encode the semantics of programs as well
  as their relational properties in trace logic, and then show how
  first-order theorem proving can be used to reason about the
  resulting trace logic formulas.  Our work is implemented in the
  tool \rapid{} and evaluated with examples coming from the
  security field.
\end{abstract}

% !TEX root = main.tex
\section{Introduction}
Program verification generally focuses on proving that all executions
of a program lie within a specified set of executions, that is, 
properties are seen as sets of traces. However, this approach is not
general enough to capture various fundamental properties, such as non-interference~\cite{GoguenM82} and
robustness~\cite{ChaudhuriGL10}. These notions are naturally modelled
as relational properties, that is as properties over sets of pairs of traces. Relational
properties are special instances of hyperproperties~\cite{ClarksonS08}, which are formally defined as sets of sets of traces.

Verification of relational properties can be achieved in different
ways.  One approach is by reduction to program verification: given a 
program $P$ and a hyperproperty $\phi$, construct a program $Q$ and a
property $\psi$, such that: (i) $Q$ verifies $\psi$ and (ii) $Q$ verifies
$\psi$ implies $P$ verifies $\phi$. The main advantage of this approach
is that (i)  can be verified using standard verification tools, whereas
(ii) is proved generically for the method used for constructing $Q$,
for instance self-composition~\cite{BartheDR04,DarvasHS05} and product
programs~\cite{BartheCK11,Churchill2019}. Another approach to verify
relational properties is to use 
relational Hoare logic~\cite{Benton04} or specialized logics that
target specific properties~\cite{AmtoftBB06}. While both approaches have
been applied successfully in several use cases, they suffer from fundamental limitations: (i) they are typically not efficient enough to scale to large programs and (ii) they are only partly automated and tailored to specific properties. %Our work addresses (ii).

\iffmcad\noindent{\bf Contributions.}\else \paragraph{\bf Contributions.}\fi
In this paper, we develop a new approach  based on reduction to
first-order reasoning, with the intent of reconciling
expressiveness and automation.

%Specifically, the contributions of this work are
%\begin{description} 
\noindent{\bf (1)} We introduce and formally characterize {\it trace logic} $\tracelogic$, an instance of many-sorted
  first-order logic with equality, which  allows
  expressing properties over
program locations, loop iterations, and computation traces (Section~\ref{sec:tracelogic}). \\[-.75em]

%\item
 \noindent{\bf  (2)} We {\it encode the semantics of programs as well as relational
    program 
  properties in $\tracelogic$} (Section~\ref{sec:tracelogic}).
  Specifically, given a program $P$ and a
relational property $F$, we construct a first-order formula $\xi$ in $\tracelogic$
such that validity of $\xi$ entails that $P$ satisfies $F$. Note that this semantic characterization stands in
contrast with methods based on product programs, Hoare logics, and 
relational Hoare logics, where verification is syntax-directed. \\[-.75em]

 \noindent{\bf (3)} We show that {\it relational properties}, such as
 non-interference, can naturally be {\it encoded in trace logic}
  (Section~\ref{sec:hyper}). \\[-.75em]

 \noindent{\bf (4)}  We {\it implemented our approach in the \rapid{}
   tool}, which relies on the first-order theorem prover
 Vampire~\cite{kovacs2013first}. 
 %for first-order logic proving.
 We conducted  experiments on
 security-relevant hyperproperties, such as non-interference and
 sensitivity. Our results show  that  \rapid{} is more expressive than
 state-of-the-art non-interference verification tools and that Vampire is better suited to the verification of security-relevant hyperproperties than state-of-the-art SMT-solvers like Z3 and CVC4.    
%\end{description}

% \begin{itemize}
% 	\item intro (2 pages)
% 	\item key ideas (2 pages)
% 	\item trace logic (16 - other = 5 pages)
% 	\item implementation (2 pages)
% 	\item experiments (2 pages)
% 	\item related work (1 page)
% 	\item Conclusion and bibliography (2)
% \end{itemize}

%!TEX root = ./main.tex

\section{Motivating Example}\label{sec:motivating}

\iffmcad
\begin{figure}  
    \begin{lstlisting}
 func main()
{
    const Int[] a;
    const Int alength;

    Int i = 0;
    Int hw = 0;

    while (i < alength)
    {
       hw = hw + a[i];
       i = i + 1;
    }
  }
\end{lstlisting}
\else
\begin{floatingfigure}[r]{40mm}
    \begin{lstlisting}[xleftmargin=.1\textwidth, xrightmargin=.53\textwidth]
 func main()
{
    const Int[] a;
    const Int alength;

    Int i = 0;
    Int hw = 0;

    while (i < alength)
    {
       hw = hw + a[i];
       i = i + 1;
    }
  }
\end{lstlisting}
\fi
\caption{Motivating example.}\vspace*{-1em}
\label{fig:running}
\iffmcad
\end{figure}
\else
\end{floatingfigure}
\fi

\noindent We motivate our work with the simple program of
Figure~\ref{fig:running}. This program  iterates over an integer-valued array
$\pv{a}$ and 
stores in the variable
$\pv{hw}$  the sum of array elements.
If $\pv{a}$  is a bitstring, then this program leaks the so-called
Hamming weight of $\pv{a}$  in the variable $\pv{hw}$.  
%Consider two arbitrary computation traces  $t_1$ and $t_2$ of that program,  of Figure~\ref{fig:running}; that is, $t_1$ and $t_2$ describe the execution of two copies of  Figure~\ref{fig:running}.
 Our aim is to prove the following {\it relational property} over two arbitrary computation traces $t_1$ and $t_2$ of Figure~\ref{fig:running}: if the elements of the array variable $\pv{a}$ in $t_1$ are component-wise equal to the elements of $\pv{a}$ in $t_2$ except for two consecutive positions $k$ and $k+1$, for some $k$, and the elements of $\pv{a}$ in $t_1$ at positions $k,k+1$ are swapped versions of the elements of $\pv{a}$ in $t_2$ (that is, the $k$-th element of $\pv{a}$ in $t_1$ is the $(k+1)$-th element of $\pv{a}$ in $t_2$ and vice-versa), then the program variable $\pv{hw}$ is the same at the end of $t_1$ and $t_2$. We formalize this property as
\begin{equation}\label{ex:running:assertion}
  \begin{array}{l}
  \iffmcad  
    \forall k_{\Int}. \Big(\big(\forall \mathit{pos}_{\Int}. ((\mathit{pos} \neql k \land \mathit{pos} \neql k+1) \rightarrow \\
     a(\mathit{pos}, t_1) \eql a(\mathit{pos},t_2)) \;\land \; a(k,t_1) \eql a(k+1,t_2) \\
\land \; a(k,t_2) \eql a(k+1,t_1) \land 0 \leq k+1<\mathit{alength} \big) \\
\qquad \rightarrow hw(\textit{end}, t_1) \eql hw(\textit{end}, t_2)\Big),
  \else  
    \forall k_{\Int}. \Big(\big(\forall \mathit{pos}_{\Int}. ((\mathit{pos} \neql k \land \mathit{pos} \neql k+1) \rightarrow a(\mathit{pos}, t_1) \eql a(\mathit{pos},t_2))\\
\qquad\land \; a(k,t_1) \eql a(k+1,t_2) \land a(k,t_2) \eql a(k+1,t_1) \land 0 \leq k+1<\mathit{alength} \big) \\
\rightarrow hw(\textit{end}, t_1) \eql hw(\textit{end}, t_2)\Big),
\fi
    \end{array}
  \end{equation}
  where $k_{\Int}$ and $\mathit{pos}_{\Int}$ respectively specify that $k$ and $\mathit{pos}$ are of sort integer $\Int$. 
  Further, $a(\mathit{pos},t_i)$ denotes the value of the element at position $\mathit{pos}$ of $\pv{a}$ in trace $t_i$,
  whereas $\textit{end}$ refers to the last program location of Figure~\ref{fig:running} (that is, line 14). 

Property ~\eqref{ex:running:assertion} is challenging to verify, since it requires theory-specific reasoning  over integers and  it involves alternation of quantifiers,
  as  the length of the array $\pv{a}$ is unbounded and the $k$-th position (corresponding to the swap) is arbitrary.
  %Proving such properties requires reasoning with both theories and quantifiers, a challenging task for state-of-the-art tools such as SMT solvers and first-order theorem provers. 
To understand the difficulty in automating such kind of reasoning, let us first illustrate how humans would naturally prove property~\eqref{ex:running:assertion}.  % come up with the following high-level proof, which - to the best of our knowledge - is the only way to prove this property.
First, split the iterations of the loop of Figure~\ref{fig:running} into three intervals: 
(i) The interval from the first iteration of the loop to the iteration where $\pv{i}$ has value $k$, 
(ii) the interval from the iteration where $\pv{i}$ has value $k$ to the iteration where $\pv{i}$ has value $k+2$, and 
(iii) the interval from the iteration where $\pv{i}$ has value $k+2$ to the last iteration of the loop. 
Next, for each of the intervals above, one proves that the equality of the value of $\pv{hw}$ in traces $t_1$ and $t_2$ is {\it preserved}; that is, if $\pv{hw}$ has the same value in $t_1$ and $t_2$ at the beginning of the interval, then $\pv{hw}$ also has the same value in $t_1$ and $t_2$ at the end of the interval.
In particular, for the first and third intervals 
% (that is, corresponding to $0\ldots k$ and $k+2,\ldots,end$),
one uses {\it inductive reasoning}, 
to conclude the \emph{preservation} of the equality across the whole interval from the step-wise preservation in the interval of the equality of the value $\pv{hw}$ in traces $t_1$ and $t_2$.
Further, for the second interval,
%, that is for $k\ldots k+2$, 
one uses commutativity of addition to prove that the value of $\pv{hw}$ in traces $t_1$ and $t_2$ is {preserved}. By combining that the values of $\pv{hw}$ in traces $t_1$ and $t_2$ are {preserved} in each of the three intervals, one finally concludes that property~\eqref{ex:running:assertion} is valid. 
%
%Third, combine the three subproofs to get a proof of the property.

While the above proof might be natural for humans, it is challenging for automated reasoners for the following reasons:
(i) one needs to express and relate different iterations in the execution of the loop in Figure~\ref{fig:running} and use these iterations to split the reasoning about loop intervals; (ii) one needs to automatically synthesize the loop intervals whose boundaries depend on values of program variables; and  (iii) one needs to  combine  theory-specific reasoning with induction for proving quantified properties, possibly with alternations of quantifiers.
In our work we address these challenges: we introduce  {\it trace
  logic},  %instance of many-sorted first-order logic
allowing us to express and automatically prove relational properties,
including property~\eqref{ex:running:assertion}. The key advantages of trace
logic are as follows. 

\noindent (i) In trace logic, program variables are encoded as unary
and binary functions over program execution timepoints. This way, we
can precisely express the value of each program variable at any
program execution timepoint, without introducing abstractions. For
Figure~\ref{fig:running}, for example, we  write $hw(\textit{end},t_1)$ to denote to the value of $\pv{hw}$ in trace $t_1$ at timepoint $\textit{end}$.
\iffmcad
\\
\fi

\noindent (ii) Trace logic further  allows arbitrary quantification
over iterations and values of program variables.
In particular, we can express and reason about iterations that depend
on (possibly non-ground)
expressions involving program variables. 
We use superposition-based first-order reasoning  to automate static analysis with trace logic and derive first-order properties about loop iterations, possibly with quantifier alternations. For Figure~\ref{fig:running}, we generate for example the property
$\exists it_{\Nat}.\big(it<n_9 \land i(l_9(it),t_1) \eql k\big),$
where $l_9$ denotes the location where the loop condition is tested
and $n_9$ denotes the first iteration of the loop upon which  the loop condition does not hold anymore.
\iffmcad
\\[-.5em]
\fi

\noindent (iii) We  guide superposition reasoning in trace logic by
using a set of lemmas statically inferred from the program semantics. These lemmas express inductive properties about the program behavior. %  about , but are independent from given programs and applicable to a wide range of programs. 
To illustrate such lemmas, we first introduce the following notation. 
%that is central for proving the first interval in our running example, can be described as follows: 
For an arbitrary program variable $\pv{v}$, let $Eq_{v}(it)$ denote that $\pv{v}$ has the same value in both traces at iteration $it$ of the loop.
For example, for every program variable $\pv{v}$ of Figure~\ref{fig:running}, we introduce the following definition: 
$$Eq_{v}(it) := v(l_9(it),t_1) \eql v(l_9(it),t_2).$$
In particular, for variable $\pv{hw}$, we introduce:
$$Eq_{hw}(it) := hw(l_9(it),t_1) \eql hw(l_9(it),t_2).$$

\noindent We then derive the following inductive lemma for each program variable $\pv{v}$: 
\begin{equation}\label{eq:running:lemma}
  \begin{array}{l}
    \forall itB_{\Nat}. 
    \iffmcad
    \Big(\big(Eq_{v}(\zero)\land \forall it_{\Nat}. ((it < itB \land Eq_{v}(it)) \\ 
    \qquad\quad \rightarrow Eq_{v}(\suc(it)))\big) \rightarrow Eq_{v}(itB)\Big),
    \else
   \Big(\big(Eq_{v}(\zero)\land \forall it_{\Nat}. ((it < itB \land Eq_{v}(it))\rightarrow Eq_{v}(\suc(it)))\big)\\
    \qquad\quad\rightarrow Eq_{v}(itB)\Big),
    \fi
     \end{array}
\end{equation}
where $itB_{\Nat}$ and $it_{\Nat}$ denote iterations $itB,it$ and $\suc(it)$ denotes the successor of $it$. 
Lemma~\eqref{eq:running:lemma} asserts that if $\pv{v}$ has the same value in traces $t_1$ and $t_2$ at the beginning of the loop (that is, at iteration $\zero$) and if the values of $\pv{v}$ are step-wise equal in traces $t_1$ and $t_2$ up to an arbitrary iteration $itB$, then the values of $\pv{v}$
are equal in traces $t_1$ and $t_2$ at iteration $itB$ (and hence the values of $\pv{v}$ are preserved in $t_1$ and $t_2$ for the entire interval up to $itB$). For Figure~\ref{fig:running}, we generate lemma~\eqref{eq:running:lemma} for $\pv{hw}$ as: 
\begin{equation}\label{eq:running:lemma:hw}
  \begin{array}{l}
    \forall itB_{\Nat}. 
    \iffmcad
      \Big(\big(Eq_{hw}(\zero)\land \forall it_{\Nat}. ((it < itB \land Eq_{hw}(it)) \\
      \qquad\quad \rightarrow Eq_{hw}(\suc(it)))\big) \rightarrow Eq_{hw}(itB)\Big).
    \else
    \Big(\big(Eq_{hw}(\zero)\land \forall it_{\Nat}. ((it < itB \land Eq_{hw}(it))\rightarrow Eq_{hw}(\suc(it)))\big)\\
    \qquad\quad\rightarrow Eq_{hw}(itB)\Big).
    \fi
 \end{array}\end{equation}
Note that lemma~\eqref{eq:running:lemma}, and in particular
lemma~\eqref{eq:running:lemma:hw} for $\pv{hw}$, is crucial for
proving that the values of $\pv{hw}$ in traces $t_1$ and $t_2$ are the
same up to iteration $k$, as considered in the relational property
of~\eqref{ex:running:assertion}. With this lemma at hand,  we
automatically prove property \eqref{ex:running:assertion} of
Figure~\ref{fig:running}, using superposition reasoning in trace logic.

\ifnotfmcad
  \vspace{1.5em}
\fi
\section{Preliminaries}
\label{sec:background}

This section fixes our terminology and programming model.

\subsection{First-order logic}
We consider standard many-sorted first-order logic with equality, where equality is denoted by $\eql$.
We allow all standard boolean connectives
and quantifiers in the language and write $s\neql t$ instead of $\neg (s \eql t)$, for two arbitrary first-order terms $s$ and $t$. 
A \emph{signature} is any finite set of symbols. We consider equality $\eql$ as part of the language; hence, $\eql$ is not a symbol. 
% The signature of a
% formula $F$ is the set of all symbols occurring in this formula. For
% example, given the function symbols $f$ and $c$, the signature of $(\forall x)(f(x) \eql c)$ is $\{f,c\}$.
We write $F_1,\ldots,F_n\vDash F$ to denote that the formula
$F_1 \land \ldots \land F_n \rightarrow F$ is a tautology.  
In particular, we write $\vDash F$, if $F$ is valid. 

By a {\it first-order theory}, or simply just {\it theory}, we  mean the set of all 
formulas valid on a class of first-order structures.
 When we discuss a theory, we call symbols
occurring in the signature of the theory {\it interpreted}, and all other
symbols {\it uninterpreted}.
In our work, we 
consider the combination (union) $\combTheory$ of the theory $\Nat$ of natural numbers and the one  $\Int$ of integers. 
The signature of $\Nat$ consists of 
standard symbols $\zero$, $\suc$, $\pred$ and $<$, respectively interpreted as {\it zero}, {\it successor},  {\it predecessor} and {\it less}. Note that $\Nat$ does not contain interpreted symbols for (arbitrary) addition and multiplication.
We use the  theory  $\Nat$ to represent and reason about loop iterations (see Section~\ref{sec:tracelogic}). 
The signature of  $\Int$  consists of the 
standard integer constants $0, 1, 2, \ldots$ and integer operators $+$, $*$ and $<$. We use the theory $\Int$ to represent and reason about integer-valued program variables (see Section~\ref{sec:tracelogic}).
Additionally we use two (uninterpreted) sorts as two sets of uninterpreted symbols:
(i) the sort Timepoint, written as $\tpsort$,  for denoting
(unique) timepoints in the execution of the program
and (ii) the sort Trace, written as $\tracesort$, for denoting
computation traces of a program.

Given a logical variable $x$ and sort $S$, 
we write $x_S$ to denote that the sort of $x$ is $S$. 
We use standard first-order interpretations/models modulo a theory
$T$, for example modulo $\combTheory$. 
We write $\vDash_{T} F$ to denote that 
$F$ holds in all models of $T$ (and hence valid).
If $I$ is a model of $T$, we write $I \vDash_{T} F$ if $F$ holds in the interpretation $I$.

% We use standard first-order interpretations/models modulo theories. 
% If $T$ is a theory, in particular $\Nat$ and $\Int$,
% we write $\vDash_{T} F$ to denote that 
% $F$ holds in all models of $T$ (and hence valid).
% If $I$ is a model of $T$, we write $I \vDash_{T} F$ if $F$ holds in the interpretation $I$.
% In the sequel we will be interested in the validity of formulas $F$ in
% the theory of {\it trace logic $\tracelogic$}. For simplicity,
% we will use $\vDash$ instead of $\vdash_{\tracelogic}$ and relativize all definitions to $\tracelogic$. 

\subsection{Programming Model $\whilelang$}

We consider programs written in a standard while-like programming language, denoted as $\whilelang$, 
with mutable and constant integer- and integer-array-variables.
The language $\whilelang$ includes standard side-effect free expressions over booleans and integers. 
Each program in $\whilelang$ consists of a single top-level function \pv{main}, 
with arbitrary nestings of \ifThenElseStatement~ and \whileStatement-statements. For simplicity,
whenever we refer to loops, we mean 
\whileStatement-loops. 
For each statement $\pv{s}$, we refer to while-statements in which $\pv{s}$ is nested in 
as \emph{enclosing loops} of $\pv{s}$. The semantics of $\whilelang$
is formalized in Section~\ref{sec:semantics:W}.

% !TEX root = main.tex
\section{Trace Logic}
\label{sec:tracelogic}

We
now introduce the concept of {\it trace logic} for expressing both the semantics 
and (relational) properties of $\whilelang$-programs.

%  which is an instantiation of first-order logic for reasoning about programs.
%
% The central idea of our language is support for describe how memory locations of the program change during the execution. In particular, we represent each program variable as a unary function, which maps each timepoint in the execution to a value, and allow full quantification over both timepoints and values.

\subsection{Locations and Timepoints}\label{sec:locations}

We consider a program in $\whilelang$ as a set of locations, 
where each location intuitively corresponds to a point in the program
at which an interpreter can stop. That is, 
for each program statement $\pv{s}$, we introduce a program location
$l_s$.
We denote by 
$l_{\mathit{end}}$ the location corresponding to the end of
the program.  

As program locations can be revisited during program executions, for
example due to the presence of loops, we model locations as follows. 
For each location $l_s$ corresponding to a program statement $\pv{s}$, 
we introduce a function symbol $l_s$ with target sort $\tpsort$ in our language, 
denoting the timepoint where the interpreter visits the location. 
For each enclosing loop of the statement $\pv{s}$, 
the function symbol $l_s$ has an argument of type $\natsort$; this
way, we distinguish between different iterations of the enclosing loop
of $\pv{s}$.
We denote the set of all such function-symbols $l_s$ as $Sig_{Tp}$.
When $\pv{s}$ is a loop, we additionally include a function symbol $n_s$ 
with target sort $\natsort$ and an argument of sort $\natsort$ for
each enclosing loop of $\pv{s}$. This way, $n_s$ denotes the iteration in which $\pv{s}$ terminates for
given iterations of the enclosing loops of $\pv{s}$. We denote the set of all such function symbols $n_s$ as $\textit{Sig}_{n}$.

\noindent\begin{example}\label{ex:locations}
Consider Figure \ref{fig:running}.
We abbreviate each statement $\pv{s}$ by the line number of the first line of $\pv{s}$. 
We use $l_6$ to refer to the timepoint corresponding to the first
assignment of $\pv{i}$ in the program. We denote by $l_{9}(\zero)$ and
$l_{9}(n_{9})$ the timepoints corresponding to evaluating the loop
condition in the first and, respectively, 
last  loop iteration. Further,   
we write $l_{11}(it)$ and $l_{11}(\suc(\zero))$ for the timepoint corresponding to the beginning of the loop body in the $it$-th and, respectively, second iteration of the loop.
Note
that $\suc(\zero)$ is a term algebra expression of $\Nat$.

\iffmcad
\qed
\fi
\end{example}

For simplicity, let us define terms over the most commonly used
timepoints. 
First, define $it^s$ to be a function, 
which returns for each while-statement $\pv{s}$ a unique variable of sort $\natsort$. 
%We use this function to consistently use variable names.
%
Second, let $\pv{s}$ be a statement, 
let $w_1,\dots,w_k$ be the enclosing loops of $\pv{s}$ 
and let $it$ be an arbitrary term of sort $\Nat$. 
\begin{align*}
	tp_\pv{s} &:= l_s(it^{w_1}, \dots, it^{w_k})&&\text{ if $\pv{s}$ is not while-statement}\\
	tp_\pv{s}(it) &:= l_s(it^{w_1}, \dots, it^{w_k}, it) &&\text{ if $\pv{s}$ is while-statement}\\
	\mathit{lastIt}_\pv{s} &:= n_s(it^{w_1}, \dots, it^{w_k})&&\text{ if $\pv{s}$ is while-statement}
\end{align*}
Third, let $\pv{s}$ be an arbitrary statement.
We refer to the timepoint where the execution of $\pv{s}$ has started (parameterized by the enclosing iterators) by
$$\mathit{start}_\pv{s} := 
\begin{cases}
	tp_\pv{s}(\zero) &\text{ if $\pv{s}$ is while-statement}\\
	tp_\pv{s} &\text{ otherwise}
\end{cases}
$$
Fourth, for an arbitrary statement $\pv{s}$, 
let $\mathit{end}_\pv{s}$ denote the timepoint which follows immediately after $\pv{s}$ has been evaluated completely (including the evaluation of substatements of $\pv{s}$):
$$
\mathit{end}_\pv{s}:=
\begin{cases}	
  \mathit{start}_{\pv{s}'}  &\text{if $\pv{s}'$ occurs after $\pv{s}$ in a context}\\
	\mathit{end}_{\pv{s}'} &\text{if $\pv{s}$ is last \iffmcad st. \else statement \fi in if-branch of $\pv{s}'$}\\
	\mathit{end}_{\pv{s}'} &\text{if $\pv{s}$ is last \iffmcad st. \else statement \fi in else-branch of $\pv{s}'$}\\
  \mathit{tp}_\pv{w}(\suc(it^{w}))\iffmcad\hspace{-0.8em}\fi &\text{if $\pv{s}$ is last \iffmcad st. \else statement \fi in body of \pv{w}} \\
  l_{\mathit{end}} &\text{otherwise}
\end{cases}
$$

\subsection{Program Variables and Expressions}\label{sec:prgVars}
In our setting, we reason about program behavior by expressing
properties over program variables $\pv{v}$. To do so, we capture the
value of program variables $\pv{v}$ at timepoints (from $\tpsort$)
in arbitary program execution traces (from $\tracesort$). Hence, we model program variables $\pv{v}$ as functions
$v: (\tpsort \times \tracesort) \mapsto \intsort$,
where $v(tp, tr)$ gives the value of $\pv{v}$ at timepoint $tp$, in trace $tr$. 
If the program variable $\pv{v}$ is an array, we add an additional argument of sort $\intsort$, 
which corresponds to the position at which the array is accessed. 
We denote by $\pvsig$ the set of such introduced function symbols denoting program
variables. 
We finally model arithmetic constants and program expressions using integer functions.

Note that our setting can be simplified for (i) non-mutable
variables -- in this case we omit the timepoint argument in the
function representation of the variable; (ii) for non-relational properties
about programs -- in this case, we only focus on one computation
trace and hence the trace argument in the function from $\pvsig$ can
be omitted. 

\begin{example}\label{ex:prgVars}
Consider again Figure \ref{fig:running}. 
By $i(l_6,tr)$ we refer to the value of program variable 
$\pv{i}$ in trace $tr$ at the moment before $\pv{i}$ is first
assigned. We 
use $\textit{alength}(tr)$ to refer to the value of variable
$\pv{alength}$ in trace $tr$. As $\pv{a}$ is unchanged in the program, we write $a(i(l_{11}(it),tr),tr)$ for 
the value of array $\pv{a}$ in trace $tr$ at position $\mathit{pos}$, 
where $\mathit{pos}$ is the value of $\pv{i}$ in trace $tr$ at timepoint $l_{11}(it)$.
In case $\pv{a}$ would have changed during the loop, we would have written $a(l_{11}(it), i(l_{11}(it), tr), tr)$ instead.
We denote by $i(l_{12}(it),tr)+1$ the value of the expression $\pv{i+1}$ in
trace $tr$ at timepoint $l_{12}(it)$.

\iffmcad
\qed
\fi
\end{example}

Consider now an arbitrary program expression $\pv{e}$.
We write $\llbracket \pv{e} \rrbracket (tp, tr)$ to denote 
the value of $\pv{e}$ at timepoint $tp$, in trace $tr$. With these
notations at hand, we introduce two definitions expressing properties
about values of expressions $\pv{e}$ at arbitrary timepoints and traces.
Consider now $v\in \pvsig$, that is a function denoting a program
variable $\pv{v}$, 
and let $tp_1, tp_2$ denote two timepoints. We define:
\iffmcad$Eq(v,tp_1,tp_2):= $\fi
\begin{equation}\label{eq:eq:expressions}
  \ifnotfmcad Eq(v,tp_1,tp_2):= \fi \left\{
		\begin{aligned}
			\forall \mathit{pos}_{\Int}. \;\;v(tp_1, \mathit{pos}, tr)& \eql v(tp_2, \mathit{pos}, tr), \hspace{-0.5em}&&\text{if \pv{v} is array}\\
			v(tp_1, tr)& \eql v(tp_2, tr), &&\text{otherwise}
		\end{aligned}
              \right.
            \end{equation}
That is, $Eq(v,tp_1,tp_2)$ in \eqref{eq:eq:expressions} states that 
the program variable $\pv{v}$ has the same values at $tp_1$ and $tp_2$. 
We also define: 
\begin{equation}\label{eq:eqall:expressions}
  \textit{EqAll}(tp_1,tp_2) := \bigwedge_{v \in \pvsig} Eq(v,tp_1,tp_2), 
\end{equation}
asserting that 
all program variables have the same values at the \iffmcad two \fi timepoints $tp_1$ and $tp_2$.

\subsection{Semantics of $\whilelang$}\label{sec:semantics:W}

We now describe the semantics of 
$\whilelang$ expressed in our {\it trace logic} $\tracelogic$. 
To do so, we state {\it trace axioms of $\tracelogic$} capturing the behavior of
possible program computation traces and then define $\tracelogic$. 

In what follows, we consider an arbitrary but fixed program $P$ in 
$\whilelang$, and give all definitions relative to $P$. 
Note that our semantics defines arbitrary executions, which are modeled by a free variable $tr$ of sort $\tracesort{}$.

\paragraph{Main-function}
Let $\pv{s\pvi{1}},\dots,\pv{s\pvi{k}}$ be statements and $P$ be a program with top-level function \while{func main \{s$_1; \ldots;$s$_k$\}}. %\pv{func main \string{s\pvi{1};$\mathtt{\dots}$;s\pvi{k}\string}}.
The semantics of $P$ is defined by the conjunction of the semantics of the statements $\pv{s\pvi{i}}$ in the top-level function 
and is the same for each trace. That is: 
\begin{equation}\label{ax:main}
  \llbracket P \rrbracket := \;\bigwedge_{i=1}^k\llbracket
  \pv{s\pvi{i}} \rrbracket.
\end{equation}

The semantics
of $P$ is then defined by structural
induction, by asserting {\it trace axioms} for each program statement
$\pv{s}$, as follows.

\paragraph{Skip}
Let $\pv{s}$ be a statement \while{skip}. The evaluation of $s$ has no effect on the value of the program variables. Hence: 
\begin{equation}\label{semantics_skip}
\llbracket \pv{s}\rrbracket := \bigwedge_{v \in \pvsig} Eq(v,\mathit{end}_\pv{s},tp_\pv{s})
\end{equation}

\paragraph{Integer assignments}
Let $\pv{s}$ be an assignment \while{v = e}, %$\pv{v = e}$
 where $\pv{v}$ is an integer program variable and
$\pv{e}$ is an expression.
We reason as follows. 
The assignment $\pv{s}$ is evaluated in one step. 
After the evaluation of $\pv{s}$, the variable $\pv{v}$ has the same value as $\pv{e}$ before the evaluation, 
and all other variables remain unchanged. Hence: 
\begin{equation}\label{semantics_int_assign}
\llbracket \pv{s}\rrbracket :=
 v(\mathit{end}_\pv{s}) \eql \llbracket \pv{e} \rrbracket(tp_\pv{s},tr) 
\land 
\hspace{-1.3em}\bigwedge_{v' \in \pvsig \setminus \{v\}} \hspace{-1em} Eq(v',\mathit{end}_\pv{s},tp_\pv{s})
\end{equation}

\paragraph{Array assignments}
Let $\pv{s}$ be an assignment \while{a[e$_1$] = e$_2$},
where $\pv{a}$ is an array variable and
$\pv{e\pvi{1}}, \pv{e\pvi{2}}$ are expressions.
We consider that the assignment is evaluated in one step. 
After the evaluation of $\pv{s}$, the array $\pv{a}$ has the same value as before the evaluation,
except for the position $\mathit{pos}$ corresponding to the value of $\pv{e\pvi{1}}$ before the evaluation,
where the array now has the value of $\pv{e\pvi{2}}$ before the evaluation. 
All other program variables remain unchanged and we have: 
\begin{subequations}
\begin{align}
\label{semantics_arr_assign_1}
\llbracket \pv{s}\rrbracket :=
&&&\forall \mathit{pos}_{\Int}. (\mathit{pos} \neql e_1(tp_\pv{s},tr) \rightarrow \nonumber\\ 
&&&a(\mathit{end}_\pv{s}, \mathit{pos}, tr) \eql a(tp_\pv{s}, \mathit{pos}, tr))\\
\label{semantics_arr_assign_2}
&\land		&&a(\mathit{end}_\pv{s}, e_1(tp_\pv{s}, tr)) \eql e_2(tp_\pv{s}, tr)\\
\label{semantics_arr_assign_3}
&\land		&&\bigwedge_{v \in \pvsig \setminus \{a\}} Eq(v,\mathit{end}_\pv{s},tp_\pv{s})
\end{align}
\end{subequations}

\paragraph{Conditional \ifThenElseStatement{} Statements}
Let $\pv{s}$ be the statement: \while{if(Cond)\{s$_1;\ldots;$s$_k$\} else \{s$_1`;\ldots;$s$_{k`}`$\}}. % $$\pv{if(Cond)\string{s\pvi{1};\dots;s\pvi{k}\string}else\string{s\pvi{1}';\dots;s\pvi{k'}'\string}}.$$
The semantics of $\pv{s}$ is defined by the following two properties: 
(i) entering the if-branch and/or entering the else-branch 
does not change the values of the variables, 
(ii) the evaluation in the branches proceeds according to the semantics 
of the statements in each of the branches.
Thus: 
%\mm{in general, avoid numbers in sentences, especially if you don't use them in the formula being described}
%
\begin{subequations}
\begin{align}
\label{semantics_ite_1}
\llbracket \pv{s} \rrbracket := 
&		    && \llbracket \pv{Cond} \rrbracket (tp_\pv{s}) \rightarrow \mathit{EqAll}(\mathit{start}_\pv{s\pvi{1}},tp_\pv{s})\\
\label{semantics_ite_2}
&\land	&\neg&\llbracket \pv{Cond} \rrbracket (tp_\pv{s}) \rightarrow \mathit{EqAll}(\mathit{start}_\pv{s\pvi{1}}',tp_\pv{s})\\
\label{semantics_ite_3}
&\land 	&& \llbracket \pv{Cond} \rrbracket (tp_\pv{s}) \rightarrow \llbracket \pv{s\pvi{1}} \rrbracket \land \dots \land \llbracket \pv{s\pvi{k}} \rrbracket\\
\label{semantics_ite_4}
&\land 	&\neg&\llbracket \pv{Cond} \rrbracket (tp_\pv{s}) \rightarrow \llbracket \pv{s\pvi{1}}' \rrbracket \land \dots \land \llbracket \pv{s\pvi{k'}}' \rrbracket
\end{align}
\end{subequations}

\paragraph{While-Loops}

Let $\pv{s}$ be the \whileStatement-statement \while{while(Cond)\{s$_1;\ldots;$s$_k$\}}.% $\pv{while(Cond)\string{s\pvi{1};\dots;s\pvi{k}\string}}$.
We refer to $\pv{Cond}$ as the \emph{loop condition}.
We use the following four properties to defined the semantics of
$\pv{s}$: 
(i) the iteration $\mathit{lastIt}_\pv{s}$ is the first iteration where the loop condition does not hold,
(ii) entering the loop body does not change the values of the variables, 
(iii) the evaluation in the body proceeds according to the semantics 
of the statements in the body, 
(iv) the values of the variables at the end of evaluating $\pv{s}$ 
are the same as the variable values 
at the loop condition location in iteration $lastIt(\pv{s})$. We then
have:

\iffmcad
$\llbracket \pv{s} \rrbracket := $
\fi %
\begin{subequations}
\begin{align}
\label{semantics_while_1}
\ifnotfmcad \llbracket \pv{s} \rrbracket := \fi      &&&\forall it^s_{\Nat}. \; (it^s<\mathit{lastIt}_\pv{s} \rightarrow \llbracket \pv{Cond} \rrbracket (tp_\pv{s}(it^s)))\\
\label{semantics_while_2}
&\land &&\neg \llbracket \pv{Cond} \rrbracket (tp(\mathit{lastIt}_\pv{s}))\\
\label{semantics_while_3}
&\land &&\forall it^s_{\Nat}. \; (it^s<\mathit{lastIt}_\pv{s} \rightarrow \mathit{EqAll}(\mathit{start}_\pv{s\pvi{1}},tp_\pv{s}(it^s))\\
\label{semantics_while_4}
&\land &&\forall it^s_{\Nat}. \; (it^s<\mathit{lastIt}_\pv{s} \rightarrow (\llbracket \pv{s\pvi{1}} \rrbracket \land \dots \land \llbracket \pv{s\pvi{k}} \rrbracket)\\
\label{semantics_while_5}
&\land && \mathit{EqAll}(\mathit{end}_\pv{s}, tp_s(\mathit{lastIt}_\pv{s}))
\end{align}
\end{subequations}

\subsection{Trace Logic $\tracelogic$} 
We now have all ingredients to define our {\it trace logic
  $\tracelogic$}, allowing us to reason about both relational and
non-relational properties of programs.

Let $\tracesig$ be a set $\{t_1,t_2,\dots\}$ of nullary function symbols of sort $\tracesort$. Intuitively, these symbols denote traces and allow us to express relational properties. 
The signature of $\tracelogic$ contains 
the symbols of the theories $\Nat$ and $\Int$ together with symbols
introduced in Section~\ref{sec:locations}-\ref{sec:prgVars}, that
is symbols denoting timepoints, 
last iterations in loops, 
program variables 
and traces.
Formally,  
% $$Sig(\tracelogic) := \qquad
% \tpsig \cup
% \lastsig \cup
% \pvsig \quad\cup \quad
% \natsig \cup 
% \intsig \quad \cup \quad
% \tracesig,$$
$$Sig(\tracelogic) \;\;:= \;\;
(\natsig \cup \intsig) 
\;\;\cup \;\;
(\tpsig \cup \lastsig \cup \pvsig \cup \tracesig)
 .$$

% \begin{definition}[$\whilelang$-Soundness]
% 	Let $p$ be a program and let $A$ be a trace logic axiom.
% 	Then $A$ is called \emph{$\whilelang$-sound}, if for any execution $E$ and for any execution-interpretation $M$ we have $M \vDash A$.
% \end{definition}

 Recall that the semantics of $\whilelang$ is defined by the trace
 axioms (7)-(11). By extending standard small-step operational semantics with
 timepoints and traces, we obtain the small-step semantics of
 $\whilelang$. For proving soundness,
  of this
 semantics, we rely on so-called 
 {\it execution-interpretation} of a program execution $E$: such an
 interpretation is a model in which for every (array) variable
 $\pv{v}$ the term $v(tp_i)$ resp. $v(tp_i,pos)$ is interpreted as the
 value of $\pv{v}$ at the execution step in $E$ corresponding to
 \ifarxiv
 timepoint $tp_i$ -- see our Appendix for more details.
 \else
 timepoint $tp_i$ -- we refer to~\cite{rapidarxiv} for more details.
 \fi
 We then introduce  $\whilelang{}$-soundness defining the  soundness
 of the semantics of $\whilelang$, as follows: 

\begin{definition}[$\whilelang$-Soundness]
  Let $p$ be a program and let $A$ be a trace logic property.
  We say that  $A$ is \emph{$\whilelang$-sound}, if for any execution-interpretation $M$ we have $M \vDash A$.
\end{definition}

By using structural induction over program statements,
we derive $\whilelang$-soundness of the semantics of
$\whilelang$. That is: 

\begin{theorem}[$\whilelang$-Soundness of Semantics of $\whilelang$]\label{thm:sound}
	For a given terminating program $p$, the trace axioms (7)-(11) are $\whilelang$-sound.
\end{theorem} 

As a consequence, the
semantics of any terminating program $p$ expressed in $\tracelogic$,
as defined in~(6), is
$\whilelang$-sound.

\subsection{Program Correctness in Trace Logic $\tracelogic$} 

Let $P$ be a program and $F$ be a first-order property of $P$, with
$F$ expressed in $\tracelogic$. We use $\tracelogic$ to express
and prove that $P$ ``satisfies'' $F$, that is $P$ is partially correct
w.r.t. $F$, as follows:
\begin{enumerate}
  \item We express $\llbracket P \rrbracket$ in $\tracelogic$, as
    discussed in Section~\ref{sec:semantics:W};
    \item We prove the partial correctness of $P$ with respect to $F$; that is,
      we prove
      $$\llbracket P \rrbracket \;\vDash_{\combTheory} \;F.$$
    \end{enumerate}

    In what follows, we first discuss (relational) properties $F$
    expressed in $\tracelogic$ (Section~\ref{sec:hyper}) and then focus on
    proving partial correctness 
    using 
    $\tracelogic$ (Section~\ref{sec:experiments}).

%!TEX root = ./main.tex
\section{Hyperproperties in Trace Logic}\label{sec:hyper}

We demonstrate the expressiveness of trace logic $\tracelogic$ by
encoding  non-interference~\cite{sabelfeld2003language} and
sensitivity~\cite{dwork2006calibrating}, two fundamental security
properties. 
\ifnotarxiv
For space restriction, we only exemplify 
non-interference and we refer to~\cite{rapidarxiv} for reasoning about
sensitivity in trace logic $\tracelogic$. 
\fi
This secition also showcases the generic lemmas,
similar to property~\eqref{eq:running:lemma}, 
introduced by our work to automate the verification of
hyperproperties. The examples considered in this section are deemed as insecure by existing syntax-driven, non-interference verification techniques, such
as~\cite{sabelfeld2003language,graf2013using}.
%Thanks to expressivity and automation of trace logic $\tracelogic$, we
%show that state-of-the-art methods are overly conservative when it comes to
%reason about such properties - namely, all examples in this section
%are proved to satisfy non-interference by our work.

\noindent\paragraph{Non-interference}
Non-interference~\cite{GoguenM82} is a security property that prevents
information flow from confidential data to public channels. It is a
so-called $2$-safety property expressing that, given two runs of a program containing high and low confidentiality variables, denoted by $H$ and $L$ respectively, if the input for all $L$ variables is the same in both runs, the output of the computation should result in the same values for $L$ variables in both traces regardless of the initial value of any $H$ variable. Intuitively, this means that no private input leaks to any public sink. 
In what follows, we let \pv{lo} denote an $L$ variable and \pv{hi} an $H$ variable.

We formalize non-interference in trace logic $\tracelogic$ as follows.
Let $l_0$ denote the first timepoint of the execution and let
$\mathit{EqTr}(v,tp)$ denote that $v$ has the same value(s) in both
traces at timepoint $tp$, that is:

\iffmcad $\mathit{EqTr}(v,tp) := $\fi
$$
\ifnotfmcad \mathit{EqTr}(v,tp) := \fi
\begin{cases}
	\forall pos_{\Int}. v(tp,pos,t_1) \eql v(tp,pos,t_2))\iffmcad\hspace{-.5em}\fi &\text{if $v$ is mutable array}\\
	\forall pos_{\Int}. v(pos,t_1) \eql v(pos,t_2)) &\text{if $v$ is constant array}\\
	v(tp,t_1) \eql v(tp,t_2)) &\text{if $v$ is mutable \iffmcad var.\else variable\fi}\\
	v(t_1) \eql v(t_2) &\text{if $v$ is constant \iffmcad var.\else variable\fi}
\end{cases}
$$

\noindent
We then express non-interference as:
\begin{equation}\label{eq:non-interference}
  (\bigwedge_{v \in L} \mathit{EqTr}(v,l_0)) \rightarrow (\bigwedge_{v \in L}
  \mathit{EqTr}(v,l_{\mathit{end}})).
 \end{equation}

\begin{example}\label{ex:noninterference3}
Consider the program illustrated in Figure~\ref{fig:noninterference3},
which branches on an $H$ guard. In the two branches, however, the $L$
variable is updated in the same way, thereby not leaking anything
about the guard. The non-interference property for this program is a
special instance of property~\eqref{eq:non-interference}, as follows:

\begin{equation}\label{eq:noninterference3:lemma}
  \mathit{EqTr}(lo,l_0) \rightarrow \mathit{EqTr}(lo,l_{\mathit{end}}).
  \end{equation}

%Note that the above property is expressed in trace logic
%$\tracelogic$.
 By adjusting superposition reasoning to trace logic $\tracelogic$ 
(see Section~\ref{sec:experiments}), we can automatically verify the
  property above. Traditional information-flow type
  systems~\cite{sabelfeld2003language} would however fail to prove this
  program secure, as they prevent any branching on $H$ guards.
  More permissive static analysis techniques based on program
  dependency graphs, such as \textsc{Joana}~\cite{graf2013using}, would
  also classify this program as insecure. %With the expressiveness of trace
%  logic, not only do we prove non-interference, and hence security, of
%  Figure~\ref{fig:noninterference3}, but we do so fully automatically
%  by generating and using trace lemmas as
%  in~\eqref{eq:noninterference3:lemma}.

  \iffmcad
  \qed
  \fi
\end{example}

\begin{figure}[tb]
\centering
    \begin{subfigure}{0.45\textwidth}
		\begin{lstlisting}
func main()
{
    const Int hi;
    Int lo;

    if(hi > 0)
    {
         lo = lo + 1;
    }
    else
    {
         lo = lo + 1;
    }
}	\end{lstlisting}
	    \caption{Branching on a \textit{high} variable.}
	    \label{fig:noninterference3}
    \end{subfigure}
    ~
	\begin{subfigure}{0.45\textwidth}
	    		\begin{lstlisting}
func main()
{
    const Int k;
    const Int lo;
    Int hi = lo;
    Int i = 0;
    Int[] output;

    while(hi < k)
    {
         output[i] = hi;
         hi = hi + 1;
         i = i + 1;
    }
}		\end{lstlisting}
	\caption{Explicit flow.}
	\label{fig:noninterference9}
	\end{subfigure}
	\caption{Examples with non-interference behaviour.}\vspace*{-1em}
	\label{fig:noninterference}
\end{figure}

Let us now focus on another interesting security example.

\begin{example}\label{ex:noninterference9}
  Figure~\ref{fig:noninterference9} models an
  interactive program  outputting on a public channel. The array
  variable $\pv{output} \in L$ models the number and content of these
  outputs, which is determined by the loop. At a first glance, this
  program might look insecure because of the explicit flow at
  $l_{11}$. Furthermore, the number of outputs, as well as their
  content, could also leak information about the secret. Indeed,
  value-insensitive  information-flow type
  systems~\cite{sabelfeld2003language} would consider this program to
  be insecure. In this specific case, however, the $H$ variable in the
  loop guard is reset with an $L$ input, and the program satisfies
  non-interference. As our semantic reasoning in trace logic
  $\tracelogic$ is value sensitive,
  our work correctly validates Figure~\ref{fig:noninterference9}. 
  proving it to be secure.  
Specifically, we prove the following property, stating that if all variables in
$L$ are equal at the beginning of the execution, then the values of
the \pv{output} array are equal after the execution:
\begin{equation}\label{eq:noninterference9:lemma}
\begin{array}{l}
	(\mathit{EqTr}(k,l_{11}) \land \mathit{EqTr}(lo,l_{11}) \land
  \mathit{EqTr}(\mathit{output},l_{11})) \\
  ~ \rightarrow 
	\mathit{EqTr}(output, l_{\mathit{end}})
% \big(\: k(t_1) = k(t_2) \:\wedge\: lo(t_1) = lo(t_2) \:\wedge\: \\ \forall pos. \; (output(l_{11}(0), pos, t_1) = output(l_{11}(0), pos, t_2)) \:\big) \rightarrow  \\ 
%  \forall pos. \; (output(l_{end}, pos, t_1) = output(l_{end}, pos, t_2)) 
      \end{array}
    \end{equation}

    \iffmcad
    \qed
    \fi
\end{example}

\ifarxiv
\noindent\paragraph{Sensitivity} Sensitivity is a property describing how much a program amplifies the distance of its inputs, which is at the core of the Laplace mechanism used to enforce differential privacy  \cite{dwork2006calibrating}.  Let the integer $k$ denote the deviation, and let $OUT$ be the set of program variables that appear in the output after the execution of the program. We can then formally define sensitivity as follows:
\begin{equation}\label{eq:sensitivity:lemma}
 \begin{array}{l}
\forall k_{\Int}, v_{\in OUT}. \: \big( |v(l_{0},t_1) - v(l_{0},t_2)| < k \\ 
\rightarrow |v(l_{end},t_1) - v(l_{end},t_2)| < k \big)
 \end{array}
\end{equation}

\begin{example}\label{example:sensitivity}
In Figure~\ref{fig:sensitivity3a}, the contents of an array $\pv{a}$
are summed up into a variable $\pv{x}$. We prove that if the values of
some variable $\pv{z}$ differ by at most $k$ between two traces while
all other array elements are equal, then the final values of $\pv{x}$
in these two traces will differ from each other by at most $k$ as
well. We express this property in trace logic $\tracelogic$ as: 
\begin{equation}\label{eq:sensitivity3a:lemma}
 \begin{array}{l}
 \forall k_{\Int}. \: \big( EqT(a,l_6) \land EqT(alength,l_6) \land |z(t_1) - z(t_2)| < k  \\
 \rightarrow |x(l_{end}, t_1) - x(l_{end}, t_2)| < k \big)
\end{array}
\end{equation}
\begin{figure}
 \begin{center}
   \begin{lstlisting}% [xleftmargin=0.35\textwidth]

func main()
{
   const Int[] a;
   const Int alength;
   const Int z;
   Int x = 0;
   Int i = 0;

   while(i < alength)
   {
        x = x + a[i];
       i = i + 1;
   }    
   x = x + z;
}   \end{lstlisting}
 \end{center}
 \caption{Example adhering sensitivity}\vspace*{-1em}
 \label{fig:sensitivity3a}
\end{figure}
\iffmcad
 \vspace{1em}
   \qed
\fi
\end{example}
\fi

Our framework generates and relies upon a set of generic \textit{trace
  lemmas} for hyperproperties, similar to lemma~\eqref{eq:running:lemma}. We now illustrate two further such lemmas.

\paragraph{Simultaneous-loop-termination\ifnotfmcad . \fi}
Our semantic formalization of $\whilelang$ in trace logic
$\tracelogic$ defines $n_s(t_1)$ to be the smallest iteration, in
which the loop condition does not hold in trace $t_1$. Due to 
well-founded orderings over naturals, there can only be one iteration
with such a property. Thus,  if we can conclude this property for any other
trace, say $t_2$, then it must be the case that
$n_s(t_2)\eql n_s(t_1)$. In our work we therefore generate and use the following trace
lemma in $\tracelogic$ (for simplicity, we omit the enclosing iterators): 

\begin{equation}\label{eq:tracelemma:termination}
  \begin{array}{l}
  \iffmcad
    \Big(\forall it. \big(it< n_s(t_1) \rightarrow \llbracket \pv{Cond} \rrbracket (l_s(it),t_2)\big) \land \\
      \neg \llbracket \pv{Cond} \rrbracket (l_s(n_s(t_1)),t_2)\Big) \rightarrow n_s(t_2)\eql n_s(t_1)
  \else
  \Big(\forall it. \big(it< n_s(t_1) \rightarrow \llbracket \pv{Cond} \rrbracket (l_s(it),t_2)\big)  \land \neg \llbracket \pv{Cond} \rrbracket (l_s(n_s(t_1)),t_2)\Big)\\
\quad \rightarrow n_s(t_2)\eql n_s(t_1)
  \fi
\end{array}
\end{equation}

\noindent Property~\eqref{eq:tracelemma:termination} is essential to
prove that the loops in both
%allows us to prove for many benchmarks that the loops in both
traces have the same last iteration, and therefore terminate after the
same number of iterations. 

\paragraph{Equality-preservation-arrays\ifnotfmcad . \fi} For an array variable
$\pv{a}$ and loop location $l$, let $Eq_{a}(\textit{it},\textit{pos})$ denote that
$\pv{a}$ at position $\textit{pos}$ has the same value in both traces at iteration
$\textit{it}$ of the loop: 
$$Eq_{a}(it,\textit{pos}) := a(l(it),\textit{pos},t_1) \eql a(l(\textit{it}),\textit{pos},t_2).$$
The following lemma over array variables is similar to the
equality-preservation-lemma~\eqref{eq:running:lemma}:
\begin{equation}\label{eq:equality-preservation-array:lemma}
\begin{array}{l}
  \forall pos_{\Int}.\forall it'_{\Nat}. \\ 
  \quad \Big(\big(Eq_{a}(\zero, pos)\land 
  \forall it_{\Nat}. ((it < it' \land Eq_{a}(it, pos)) \\
  \qquad \quad\rightarrow Eq_{a}(\suc(it),pos))\big)\rightarrow Eq_{a}(it',pos)\Big)
\end{array}
\end{equation}

We conclude by emphasizing that trace lemmas,
such as~\eqref{eq:tracelemma:termination} and~\eqref{eq:equality-preservation-array:lemma}, are expressed in
trace logic $\tracelogic$ and automatically
generated by our approach. %Such, and similar trace lemmas, are used
\section{Implementation and Experiments} \label{sec:experiments}
%This section describes our implementation and reports on our experiments for proving relational properties.

% \paragraph{Discussion on Formalization}
% A big part of our work was to optimize our formalization, 
% so that the resulting reasoning task is as simple as possible 
% for superposition-based provers.
%
% \begin{itemize}
% 	\item We are able to refer to loop iterations beyond the last iteration. 
% 	We do not need to reason about those iterations 
% 	since they are not part of any computation, 
% 	but include them nonetheless in order to achieve a simple encoding.
% 	\item Our encoding defines values both in the iterations beyond the last iteration of a loop 
% 	and in the non-visited branch of an \ifThenElseStatement statement.
% 	This over-definition yields a simpler and more efficient encoding. 
% 	In particular, we generate in many cases a unit equality clause instead of a clause containing
% 	an inequality and an equality, and as a result the proof search is more effective.
% 	\item We use different sorts for values of program variables and timepoints, 
% 	which has the main advantage that we are able to control 
% 	which connections the prover tries to establish. 
% 	In particular, the solver does not get lost in reasoning 
% 	about addition or multiplication for timepoints.
% 	\item Timepoints are independent from traces.
% \end{itemize}

\subsection{Implementation}
We implemented our approach in the tool \rapid{}\footnote{\url{https://github.com/gleiss/rapid}},
which consists of nearly 13,000 lines of C++ code.
\rapid{} takes as input a program written in $\whilelang$ and a property expressed in trace logic $\tracelogic$. 
It then generates axioms written in trace logic $\tracelogic$ corresponding to the semantics of the program and outputs both the axioms and the property in the \smtlib{} syntax~\cite{barrett2017smtlib}. The produced \smtlib{} encoding is further passed within \rapid{} to the first-order theorem prover \vampire{} for proving validity of the property (i.e. partial correctness). \vampire{} searches for a refutation of the desired property by saturating the provided encoding with respect to a set of inference rules such as resolution and superposition~\cite{kovacs2013first}. 

\paragraph{Inductive Reasoning\ifnotfmcad . \fi}
Trace logic $\tracelogic$ encodes loop-iterations using counters of sort $\Nat$. Hence, there are consequences of the semantics which can only be derived using inductive reasoning. Automating induction is however challenging: state-of-the-art SMT solvers and theorem-provers are not able to automatically infer and prove most (inductive) consequences needed  by \rapid{}. In order to address this problem, (i) we identified some of the most important applications of induction that are useful for many programs and (ii)  formulated the corresponding inductive properties in trace logic as \emph{trace lemmas}. Some of these lemmas are described in Section~\ref{sec:motivating} and Section~\ref{sec:hyper}. Each trace lemma is logically implied by standard induction axioms of natural numbers and the semantics of the program. \rapid{} generates trace lemmas for each variable and each loop of the program and adds them as axioms to its \smtlib{} output.

\paragraph{Theory Reasoning}
Reasoning with theories in the presence of quantifiers is yet another challenge for automated reasoners, and hence for \vampire{}. Different theory encodings lead to very different results. 
In \rapid{}, we model integers using the built-in support for integers in \vampire{}. We experimented with various sound but incomplete axiomatization of integers. We used \vampire{} with all its built-in theory axioms  (option \texttt{-tha on}, default), as well as with a partial, but most relevant set of theory axioms (option \texttt{-tha some}) which we extended with specific integer theory axioms. 
% These axioms have now been added automatically by \rapid{} to the \smtlib{} encoding of our benchmarks. We leave the task of adding these axioms as built-in theory axioms of \vampire{} as an interesting line of future work.
%
Natural numbers are modeled in \rapid{} as a term algebra $(\zero,\suc, \pred)$, for which efficient reasoning engines already exist~\cite{kovacs2017coming}. In order to express the ordering of natural numbers, we manually add the symbol $<$, together with an (incomplete) axiomatization.
In \rapid{}, we also  experimented with clause splitting by calling \vampire{} both with and without its \avatar{} framework \cite{voronkov2014avatar} (options \texttt{-av on/off}, with {\tt on} as default).

\subsection{Benchmarks and Experimental Results}\label{sec:experiments:benchmarks}

%We evaluated \rapid{} on a set of 27 benchmarks expressing hyperproperties.
To compensate the lack of general benchmarks for first-order hyperproperties, 
we collected a set of  27 verification problems for evaluating our work in \rapid{}.
Our benchmarks describe $2$-safety properties relevant in the security domain,  such as non-interference and sensitivity. The individual benchmark programs consist of up to 50 lines of code each. 

\rapid{} produced the \smtlib{}-encodings for each benchmark in less than a second. These encodings were passed to \vampire{},  as well as to the SMT solvers \textsc{Z3}~\cite{Z3} and \textsc{CVC4}~\cite{CVC4} for comparison purposes, to establish the correctness of the input property.
We ran each prover
with a 60 seconds time limit. All experiments were carried out on an Intel
%\textsuperscript{\textregistered{}} 
Core
%\texttrademark{} 
i5 3.1Ghz machine with 16 GB of RAM.

Our experimental results are summarized in Table~\ref{tab:results}. The first four columns report on results by running \vampire{} on the \rapid{} output. 
The columns denoted with \texttt{S/F} refer to \vampire{} options for partial/full theory reasoning (option \texttt{-tha some/on}) respectively. \texttt{A} refers to the use of the \avatar{} (option \texttt{-av on}) in conjunction with one of the theory options, hence  columns \texttt{S+A} and \texttt{F+A}. 
The last two columns of Table~\ref{tab:results} summarize our results of running \textsc{Z3} and \textsc{CVC4} on the \rapid{} output. 
The rows denoted \textit{Total \vampire{}} and \textit{Unique \vampire{}} sum up the total and unique numbers of examples proven with the setting of the corresponding column. \iffmcad Example \else The example\fi~{\tt 4-hw-swap-in-array} in Table~\ref{tab:results} is our running example from Figure~\ref{fig:running}, whereas \iffmcad the \fi benchmarks~{\tt 3-ni-high-guard-equal-branches} and {\tt 9-ni-equal-output} correspond to Figure~\ref{fig:noninterference3} and Figure~\ref{fig:noninterference9},  respectively. 

% As a first experiment, we were interested in the number of \rapid{} benchmarks \vampire{}~ proves using at least one of  the four considered proving strategies.
\vampire{} proved 25  \rapid{} encodings out of the 27 benchmark problems.
 Table~\ref{tab:results} shows that the option \texttt{S+A} seems to be the most successful, with four unique benchmarks proven. %However, comparing full and partial reasoning also shows that both options work well for different sets of problems. 
 While two of our benchmarks were not proven by \vampire{} with our current set of automatically generated \rapid{} lemmas, these problems could actually be proved by \vampire{} by using only a subset of trace lemmas, i.e. by removing unnecessary lemmas manually.
 % This accelerated proof search to find the right clauses faster before a blow-up due to theory reasoning prevents \vampire{} from finding a proof in the given time limit.
Improving theory reasoning in \vampire{}, and in general in superposition proving, would further improve the efficiency of \rapid{}. In particular, designing better reasoning support for transitive relations like $<_{\Nat}$ and $<_{\Int}$ is an  interesting further line of research. 

	\begin{table}[tb]\footnotesize\centering
		\iffmcad\setlength{\tabcolsep}{4pt}\else \setlength{\tabcolsep}{8pt}\fi
		\begin{tabular}{|c|c|c|c|c|c|c|}
			\hline
			\multirow{2}{*}{Benchmarks}                          & \multicolumn{4}{c|}{Vampire}                                                                          & \multirow{2}{*}{CVC4}   & \multirow{2}{*}{Z3}    \\\cline{2-5}
			& S                       & S+A & F                       & F+A                   &                         &                         \\ \hline
			1-hw-equal-arrays                    & $\checkmark$            & $\checkmark$            & -                       & $\checkmark$            & $\checkmark$            & $\checkmark$            \\ \hline
			2-hw-last-position-swapped                    & -                       & $\checkmark$            & -                       & -                       & $\checkmark$            & $\checkmark$            \\ \hline
			3-hw-swap-and-two-arrays                    & -                       & $\checkmark$            & -                       & -                       & -                       & -                       \\ \hline
			4-hw-swap-in-array-lemma              & -                       & $\checkmark$            & -                       & -                       & -                       & -                       \\ \hline
			4-hw-swap-in-array-full               & -            & $\checkmark$            & -                       & -                       & -                       & -                       \\ \hline
			1-ni-assign-to-high                  & $\checkmark$            & $\checkmark$            & $\checkmark$            & $\checkmark$            & $\checkmark$            & $\checkmark$            \\ \hline
			2-ni-branch-on-high-twice                  & $\checkmark$            & $\checkmark$            & $\checkmark$            & $\checkmark$            & $\checkmark$            & $\checkmark$            \\ \hline
			3-ni-high-guard-equal-branches                  & $\checkmark$            & $\checkmark$            & $\checkmark$            & $\checkmark$            & $\checkmark$            & $\checkmark$            \\ \hline
			4-ni-branch-on-high-twice-prop2                  & $\checkmark$            & $\checkmark$            & -                       & -                       & $\checkmark$            & $\checkmark$            \\ \hline
			5-ni-temp-impl-flow                  & -                       & -                       & $\checkmark$            & $\checkmark$            & $\checkmark$            & $\checkmark$            \\ \hline
			6-ni-branch-assign-equal-val                  & -                       & -                       & $\checkmark$            & $\checkmark$            & $\checkmark$            & $\checkmark$            \\ \hline
			7-ni-explicit-flow                  & $\checkmark$            & $\checkmark$            & $\checkmark$            & $\checkmark$            & $\checkmark$            & $\checkmark$            \\ \hline
			8-ni-explicit-flow-while                  & $\checkmark$            & $\checkmark$            & -                       & $\checkmark$            & $\checkmark$            & $\checkmark$            \\ \hline
			9-ni-equal-output                  & $\checkmark$            & -                       & -                       & -                       & -                       & $\checkmark$            \\ \hline
			10-ni-rsa-exponentiation                 & $\checkmark$            & $\checkmark$            & $\checkmark$            & $\checkmark$            & $\checkmark$            & -                       \\ \hline
			1-sens-equal-sums                       & $\checkmark$            & $\checkmark$            & $\checkmark$            & $\checkmark$            & $\checkmark$            & $\checkmark$            \\ \hline
			2-sens-equal-sums-two-arrays                       & $\checkmark$            & $\checkmark$            & $\checkmark$            & $\checkmark$            & -                       & -                       \\ \hline
			3-sens-abs-diff-up-to-k                       & -                       & -                       & -                       & -                       & $\checkmark$            & $\checkmark$            \\ \hline
			4-sens-abs-diff-up-to-k-two-arrays                       & -                       & -                       & -                       & -                       & -                       & -                       \\ \hline
			5-sens-two-arrays-equal-k                       & $\checkmark$            & $\checkmark$            & $\checkmark$            & $\checkmark$            & -                       & -                       \\ \hline
			6-sens-diff-up-to-explicit-k                       & $\checkmark$            & $\checkmark$            & $\checkmark$            & $\checkmark$            & -                       & -                       \\ \hline
			7-sens-diff-up-to-explicit-k-sum                       & -                       & -                       & $\checkmark$            & $\checkmark$            & -                       & -                       \\ \hline
			8-sens-explicit-swap                       & -                       & -                       & $\checkmark$            & $\checkmark$            & -                       & -                       \\ \hline
			9-sens-explicit-swap-prop2                       & -                       &                         & $\checkmark$            & $\checkmark$            & -                       & -                       \\ \hline
			10-sens-equal-k                      & $\checkmark$            & $\checkmark$            & $\checkmark$            & $\checkmark$            & -                       & -                       \\ \hline
			11-sens-equal-k-twice                      & $\checkmark$            & $\checkmark$            & $\checkmark$            & $\checkmark$            & -                       & -                       \\ \hline
			12-sens-diff-up-to-forall-k                      & -                       & -                       & $\checkmark$            & $\checkmark$            & $\checkmark$            & -                       \\ \hline
			Total \vampire  & \multicolumn{1}{c|}{15} & \multicolumn{1}{c|}{18} & \multicolumn{1}{c|}{17} & \multicolumn{1}{c|}{19} &                         &                         \\ \hline
			Unique \vampire & \multicolumn{1}{c|}{1}  & \multicolumn{1}{c|}{4}  & \multicolumn{1}{c|}{0}  & \multicolumn{1}{c|}{0}  & \multicolumn{1}{c|}{}   & \multicolumn{1}{c|}{}   \\ \hline
			Total                              & \multicolumn{4}{c|}{25}                                                                               & \multicolumn{1}{c|}{14} & \multicolumn{1}{c|}{13} \\ \hline
		\end{tabular}	
              \caption{\rapid{} results with \vampire{}, \textsc{Z3} and \textsc{CVC4}.\vspace*{-1em}}
	\label{tab:results}
	\end{table}

We also compared the performance of \vampire{} on the \rapid{} examples to the performance of \textsc{Z3} and \textsc{CVC4}. 
Unlike \vampire{},  \textsc{Z3} and \textsc{CVC4} proved only 13 and 14 examples, respectively. Our results thus showcase that superposition reasoning, in particular \vampire{}, is better suited for proving first-order hyperproperties, as many of these properties involve heavy use of quantifiers, including alternations of quantifiers (such as for 
example~{\tt 4-hw-swap-in-array}  corresponding to Figure~\ref{fig:running}). Moreover, \rapid{} proved security of examples that were classified insecure by existing techniques~\cite{graf2013using,sabelfeld2003language},  such as {\tt 3-ni-high-guard-equal-branches} and {\tt 9-ni-equal-output}.

\ifnotfmcad
  \vspace{-1.5em}
\fi

%!TEX root = ./main.tex
\section{Related Work}\label{sec:related}
%We compare our work with two line of research:
%(i) deductive verification approaches and (ii) verification of
%security properties.

\noindent{\bf Deductive verification}.  
Most verification approaches use a state-based language 
to express programs and properties about them, 
and use invariants to establish program correctness~\cite{bjorner2015horn}. 
% As described in Cousot95, one can see state-based representations as abstractions of trace-based representations. The abstraction sacrifies the ability to directly express timepoints including the possibility to express dependencies of timepoints on values of program variables, in order to reason in a simpler language. 
% In particular, the abstracted representation contains much less quantification, which makes state-based languages targetable for smt-solvers. 
%
Such invariants loosely correspond to a fragment of trace logic, 
where formulas only feature universal quantification over time -- but
no existential quantification. 
The lack of existential, and thus alternating, quantification 
makes these works suitable for automation via
SMT-solving~\cite{hoder2012generalized,Gurfinkel16} and hence
applicable for programs where full first-order logic is not needed, for
instance programs involving mainly integer variables and function
calls. %
For program properties expressed in full first-order
logic, such as over unbounded arrays,  
existing methods are  yet not able to automatically verify
program correctness. 
We argue that the missing expressiveness is the problem here, 
since one usually needs to be able to express 
arbitrary dependencies of timepoints and values,  
if custom code is used to iterate through an array 
or more generally through a data structure. Our trace logic supports
such kind of first-order reasoning. 

% One particular instance of state-based verification, which has shown to be very effective, generates invariants by generalizing from bounded executions \tood{cite pdr, Z3-fixedpoint-engine and spacer}.
% Extending the existing approaches to handle trace-based reasoning seems to be very difficult.
% While infering quantified invariants from examples is already very hard, since there are many choices to generalize the bounded executions, 
% it seems to be much harder to generalize from examples to invariants containing quantifier-alternations,
% since the space of potential invariants is even bigger.

Our approach to automate induction using trace lemmas is related to 
template-based invariant generation methods~\cite{colon2003linear,gupta2009invgen}. 
%It remains however unclear how to define a set of general templates, which are applicable for many programs. 
%
%Thanks to the expressiveness of trace logic,
Our trace lemmas are
however more expressive  than existing templates and we %Further,  %Our trace lemmas are not With our trace lemmas 
%we argue that potential limitations of template-based methods don't necessarily carry over to our approach: 
%Since our language is much more expressive, we are able to formulate very general lemmas, 
%by using superposition-based reasoning over our trace lemmas,
%we
automatically derive trace lemmas. % similar to templates. 

Program analysis by first-order reasoning is also studied
in~\cite{LPAR-22:Loop_Analysis_by_Quantification}, where
program semantics  is expressed in extensions of Hoare Logic with
explicit timepoints. 
Unlike~\cite{LPAR-22:Loop_Analysis_by_Quantification}, we do not rely
on 
%
%In contrast to their work, 
%which formalizes the semantics of programs by extending Hoare Logic with explicit timepoints, 
%our approach doesn't need
an intermediate program (Hoare) logic, but reason also about
relational 
properties. While~\cite{LPAR-22:Loop_Analysis_by_Quantification} can only
handle simple loops, our work supports  a standard
while-language with explicit locations and arbitrary nestings of
statements.

%Another exploration of
First-order reasoning for program analysus is also addressed in~\cite{beckert2001}, by introducing dynamic trace logic: an extension of dynamic logic  with modalities for reasoning about traces.
A custom sequent calculus is proposed in~\cite{beckert2001}, implying that automating the work would require the design of specialised sequent calculus provers. Unlike~\cite{beckert2001}, our work is fully automated. % Since our work is an instance of standard first-order logic we can apply state-of-the-art theorem provers directly.
Further, our work preserves the control-flow structure of programs by introducing function symbols and automates inductive reasoning using trace lemmas. % Unlike~\cite{beckert2001}, our work is fully automated. %Aditionally, we experimentally showcase our work, which is not the case in \cite{beckert2001}.

%\LK{do we need this? very similar to~\cite{ClarksonFKMRS14}}
%In~\cite{emmer2012epr}, first-order reasoning for bounded model checking is studied. Yet, this work is restricted to the EPR fragment, whereas our semantics exploits full first-order logic. In particular, we rely on non-constant function symbols to model arrays and timepoints; such functions cannot be used in EPR.

\noindent{\bf Relational verification. } Verification of relational- and hyperproperties is an active area of
research, with applications in programming languages and compilers,
security and privacy; see~\cite{BeckertU18} for an overview. Various static analysis techniques have been proposed to analyze non-interference, such as type systems~\cite{sabelfeld2003language} and graph dependency analysis~\cite{graf2013using}. Type systems proved also effective in the verification of privacy properties for cryptographic protocols~\cite{Eigner:2013,Barthe:2014,Cortier:2015,Cortier:2017:TSP,Grimm:2018:POST}.    Relational Hoare logic was introduced in~\cite{Benton04}
and further extended in~\cite{BartheCK11,BartheCK13} for defining product
programs to reduce relational verification to standard verification.
All these works closely tie verification to the syntactic program structure, thus limiting their applicability and expressiveness.
As already argued, our work allows proving security of examples that were so far classified as insecure by some of the aforementioned methods~\cite{graf2013using,sabelfeld2003language}.  Recently, \cite{Grimm:2018:MFR} encodes relational properties through refinement types in F*~\cite{FStar}. While still being syntax driven, \cite{Grimm:2018:MFR} can potentially verify  semantic properties by using SMT solving, although this typically requires the manual insertion and proof of program-dependent lemmas, which is not the case for us. 

In~\cite{GodlinS13} bounded model checking is proposed for program
equivalence. Program equivalence is reduced in~\cite{FelsingGKRU14} 
to proving a set of Horn clauses, by combining a relational
weakest precondition calculus with SMT-based reasoning.
However, when addressing programs with different control flow as
in~\cite{FelsingGKRU14}, user guidance is required for proving program
equivalence. Program equivalence is also studied
in~\cite{ZhouHH17,KwonHE17} for proving information flow properties. Unlike these works, we are not limited to
SMT solving but automate the verification of relational properties expressed in full
first-order theories, possibly with alternations of quantifiers.

Motivated by applications to translation
validation, the work of~\cite{NamjoshiS16} develops powerful
techniques for proving correctness of loop transformations. Relational methods for reasoning about program
versions and semantic differences are also introduced in~\cite{PartushY14,LahiriHKR12}. Going
beyond relational properties, an SMT-based framework for verifying
$k$-safety properties is introduced in~\cite{SousaD16} and further
extended~\cite{SousaDL18} for proving
correctness of 3-way merge.  While these works focus
on high-level languages, many others consider low-level languages,
see~\cite{SmithD08,SteppTL11,SSCA13,BalliuDG14} for some exemplary
approaches. Further afield, several authors have introduced logics for
modelling hyperproperties. Unlike these
works, trace logic allows expressing first-order relational properties
and automates reasoning about such properties by first-order theorem
proving, overcoming thus the SMT-based limitations of quantified reasoning.

Finally, in~\cite{ClarksonFKMRS14} HyperLTL and HyperCTL$\mbox{}^*$ is introduced to model temporal and relational properties properties. However, these logics support only decidable fragments of first-order logic and thus cannot handle relational properties  with non-constant function symbols. As such, security and privacy properties over unbounded data structures/uninterpreted functions cannot be encoded or ve\-ri\-fied.

\section{Conclusion}\label{sec:concl}

We introduced trace logic for automating the verification of
relational program  properties of  imperative programs. We showed that
program semantics as well as relational properties can naturally be
encoded in trace logic as first-order properties over 
program locations, loop iterations and computation traces. We combined
trace logic with 
superposition proving and implemented our work in the \rapid{}
tool. While our work already outperforms SMT-based approaches,
%solver% current experiments demonstrate the efficiency and automation of our
%approach, outperforming SMT solvers,
we are convinced that improving superposition reasoning with
both theories and quantifiers would further strengthen the use of
trace logic for relational verification. %We leave this challenge as an
%interesting line of future work.

\subsection*{Acknowledgements.} This work was funded by
the ERC Starting Grant 2014 SYMCAR 639270,
the ERC Proof of Concept Grant 2018 
SYMELS 842066,
the Wallenberg Academy Fellowship 2014 TheProSE, 
the Austrian FWF research projects W1255-N23 and RiSE S11409-N23, the ERC Consolidator Grant 2018 BROWSEC 771527,  by the Netidee  projects EtherTrust  2158 and PROFET  P31621,  and by the FFG projects PR4DLT  13808694 and  COMET K1 SBA.

%\newpage
\iffmcad
\bibliographystyle{IEEEtran}
  \ifnotarxiv  
  \IEEEtriggeratref{24}
  \fi
  \bibliography{IEEEabrv,bib}
\else
  \bibliographystyle{abbrv}
  \bibliography{bib.bib}
\fi

\ifarxiv
%!TEX root = ./main.tex

\appendix

\ifnotfmcad
\section{Appendix}
\fi

\subsection{Small-step operational semantics of $\mathcal{W}$}
\label{sec:semantics}
In this subsection, we recall standard definitions from small-step operational semantics.
\begin{definition}
	Let $p$ be a program. Then a \emph{state} $\sigma$ is a function which (i) maps each integer-variable $\pv{v}$ of $p$ to a concrete value $\sigma(\pv{v}) \in \Int$ and (ii) maps each array-variable $\pv{v}$ and each value $pos \in \Int$ to a value $\sigma(\pv{v},pos) \in \Int$.
\end{definition}

\begin{definition}
	A \emph{configuration} is a pair $\langle p, \sigma \rangle$, where we refer to $p$ as the \emph{continuation} and $\sigma$ is a state.
\end{definition}

The execution of a single step in the program is defined by the rules of Figure~\ref{fig:operational-semantics}.
Our presentation is semantically equivalent to standard small-step operational semantics, but differs syntactically in three points, in order to simplify later definitions and theorems: 
(i) program-expressions are evaluated on the fly without introducing explicit steps
(ii) the relation between the state $\sigma$ in the original configuration and the state $\sigma'$ in the resulting configuration is explicitly described using a formula (in contrast to using the same variable $\sigma$ twice) and 
(iii) we annotate while-statements with counters to ensure the uniqueness of continuations during the execution, see Section~\ref{subsec:uniqueness}.

\begin{figure}[tb]
	\centering
		\begin{prooftree}
			\AxiomC{$\sigma' = \sigma$}
			\LeftLabel{[\textit{skip}]}
			\UnaryInfC{$\langle \skipOp{};p,\sigma\rangle \weirdarrow{} \langle p,\sigma'\rangle$}
			\end{prooftree}
			
			\begin{prooftree}
			\AxiomC{$\sigma' = \sigma[v \mapsto \llbracket e \rrbracket(\sigma)]$}
			\LeftLabel{[\textit{asg}]}
			\UnaryInfC{$\langle \texttt{v} := \texttt{e};p,\sigma\rangle \weirdarrow{} \langle p,\sigma'\rangle$}
			\end{prooftree}
			
			\begin{prooftree}
			\AxiomC{$ \llbracket c \rrbracket (\sigma) = \textit{true} $}
			\AxiomC{$\sigma' = \sigma$}
			\LeftLabel{$[\textit{ite}_{\textit{T}}]$}
			\BinaryInfC{\stackanchor{\langle \iteOp{c}{p_1}{p_2};p,\sigma\rangle \weirdarrow{}} {\langle p_1;p,\sigma'\rangle}}
			\end{prooftree}
			
			\begin{prooftree}
			\AxiomC{$ \llbracket c \rrbracket (\sigma) = \textit{false} $}
			\AxiomC{$\sigma' = \sigma$}
			\LeftLabel{$[\textit{ite}_{\textit{F}}]$}
			\BinaryInfC{\stackanchor{\langle \iteOp{c}{p_1}{p_2};p,\sigma\rangle \weirdarrow{}} {\langle p_2;p,\sigma'\rangle}}
			\end{prooftree}
			
			\begin{prooftree}
			\AxiomC{$ \llbracket c \rrbracket (\sigma) = \textit{true} $}
			\AxiomC{$\sigma' = \sigma$}
			\LeftLabel{[\textit{while}$_T$]}
			\BinaryInfC{\Shortunderstack[c]{{\langle \whileOp[i]{c}{p_1};p,\sigma\rangle \weirdarrow{}} {\langle p_1; \whileOp[i+1]{c}{p_1};p,\sigma'\rangle}}}
			\end{prooftree}
			
			\begin{prooftree}
			\AxiomC{$ \llbracket c \rrbracket (\sigma) = \textit{false} $}
			\AxiomC{$\sigma' = \sigma$}
			\LeftLabel{[\textit{while}$_F$]}
			\BinaryInfC{$\langle \whileOp[i]{c}{p_1};p,\sigma\rangle \weirdarrow{} \langle p,\sigma'\rangle$}
			\end{prooftree}
	\caption{Small-step operational semantics of $\whilelang$}
	\label{fig:operational-semantics}
\end{figure}

A program is executed by iteratively transforming the initial configuration according to the rules of Figure~\ref{fig:operational-semantics} until the continuation becomes $end$. We annotate each while-statement in the initial configuration of the execution with counter $0$:
\begin{definition}
	Let $p$ be a program, let $p'$ be the result of annotating each while-loop in $p$ with counter $\zero$ and let $\sigma$ be an arbitrary state. Then $\langle p', \sigma\rangle$ is called \emph{initial configuration}.
\end{definition}

\begin{definition}
	Let $p$ be a program and $C,C_1,C_2$ be configurations. A \emph{partial execution from $C_1$ to $C_2$} is a derivation in the inference system of small-step operational semantics starting at $C_1$ and ending in $C_2$. An \emph{execution of $p$} is a partial execution from an initial configuration to a configuration $\langle end, \sigma \rangle$ for an arbitrary state $\sigma$. If there exists a partial execution starting at the initial configuration and ending in $C$, we say that $C$ is \emph{reachable}.
\end{definition}

\subsection{Separating subprograms and state}
\label{subsec:uniqueness}
Our presentation of operational semantics features counters. We now show that as a result, if $r_1$ and $r_2$ are continuations occuring in the same execution, then $r_1$ and $r_2$ are different. This implies that we do not need to know about the state to distinguish different configurations and allows us to separate the continuation from the state.

\begin{theorem}[Uniqueness]
	Let $p$ be a program and let $\langle r_1,\sigma_1 \rangle$ and $\langle r_2,\sigma_2\rangle$ be configurations occuring in the execution of $p$. Then $r_1 \neq r_2$.
\end{theorem}

\iffmcad
\begin{IEEEproof}
\else
\begin{proof}
\fi
	Let $p,p_1,p_2$ be subprograms, let $s$ be a single statement and let $C$ be a condition. Consider the minimal relation $>_1$ which satisfies the following conditions and consider its transitive closure $>$.
	\begin{align*}
		s;p &\;>_1\;p\\
		\iteOp{C}{p_1}{p_2};p & \;>_1\; p_1;p\\
		\iteOp{C}{p_1}{p_2};p & \;>_1\; p_2;p\\
		\whileOp[i]{C}{p_1};p & \;>_1\; \\
		&\hspace{-2em}p_1;\whileOp[i+1]{C}{p_1};p
	\end{align*}
	It is an easy exercise to establish that $>$ is a strict partial order on continuations.
	Next, $skip$, $asg$, $whileF$ and $end$ reduce the ordering according to the first condition, $iteT$ and $iteF$ reduce the ordering according to the second resp. third condition and $whileT$ reduce the ordering according to the fourth condition. In particular, we are able to conclude $r_1>r_2$, which immediately implies $r_1 \neq r_2$ due to the irreflexivity of $>$.
\iffmcad
\end{IEEEproof}
\else
\end{proof}
\fi
Having established the uniqueness, we are now able to speak of \emph{the} state at a given continuation $p$ (and annotate it as $\sigma(p)$). As a result, a configuration is fully described by the continuation. We therefore omit the state $\sigma(p)$ in any configuration $\langle p,\sigma(p) \rangle$ and write $\langle p \rangle$ instead. 
Finally we use the fact that we have finitely many program variables $v_1,\dots,v_n$, and split up $\sigma(p)$ into $\sigma_{v_1}(p),\dots, \sigma_{v_n}(p)$, which we simply write as $v_1(p),\dots, v_n(p)$. 

\subsection{Mapping timepoints to continuations}
Small-step operational semantics describes only the next step in an execution, whereas structural semantics, and trace logic semantics in particular, also describes the complete execution of each 
substatement.

Recall that the definitions of  $\mathit{start}_s$ and $\mathit{end}_s$ from Section~\ref{sec:locations} describe the timepoints of the start, respectively end of a partial execution of a statement $s$. To connect the two worlds of operational and structural semantics we provide a mapping $R$ from such timepoints to continuations:

\begin{definition}
	Let $R$ be 
	\begin{align*}
		&&R(tp_s) &:= s;R(\mathit{end}_s) &&\text{ if $s$ is non-loop}&&\\
		&&R(tp_s(it)) &:= s^{it};R(\mathit{end}_s) &&\text{ if $s$ is loop}&&
	\end{align*} 
\end{definition}

We are now able to describe configurations using $tp_s$, $\mathit{start}_s$ and $\mathit{end}_s$. In particular, we are able to instantiate each rule to a new rule, whose configurations can be described using $tp_s$, $\mathit{start}_s$ and $\mathit{end}_s$.  The instantiated rules produce the same reachable configurations as the original rules, and are presented in Figure~\ref{fig:operational-semantics-timepoints}.

\begin{figure}[tb]
	Let $s$ be a skip-statement. Instantiating the $skip$-rule with $p := R(\mathit{end}_s)$ yields
	\begin{prooftree}
		\AxiomC{$\sigma' = \sigma$}
		\LeftLabel{[\textit{skip}]}
		\UnaryInfC{$\langle R(\mathit{start}_{s}),\sigma\rangle \weirdarrow{} \langle R(\mathit{end}_{s}),\sigma'\rangle$}
	\end{prooftree}
	Let $s$ be an assignment \while{v = e}. Instantiating the $asg$-rule with $p := R(\mathit{end}_s)$ yields	
	\begin{prooftree}
		\AxiomC{$\sigma' = \sigma[v \mapsto \llbracket e \rrbracket(\sigma)]$}
		\LeftLabel{[\textit{asg}]}
		\UnaryInfC{$\langle R(\mathit{start}_{s}),\sigma\rangle \weirdarrow{} \langle R(\mathit{end}_{s}),\sigma'\rangle$}
	\end{prooftree}
	
	Let $p_1$ and $p_2$ be \while{s$_1;\ldots;$s$_k$} resp. \while{s'$_1;\ldots;$s'$ _{k`}$} and let $\pv{s}$ be $\texttt{if}(Cond)\,\{p_1\}\, else \,\{p_2\}$. Instantiating the rules $ite_T$ and $ite_F$ with $p := R(\mathit{end}_s)$ yields the two rules
	\begin{prooftree}
		\AxiomC{$ \llbracket c \rrbracket (\sigma) = \textit{true} $}
		\AxiomC{$\sigma' = \sigma$}
		\LeftLabel{$[\textit{ite}_{\textit{T}}]$}
		\BinaryInfC{$\langle R(\mathit{start}_{s}),\sigma\rangle \weirdarrow{}\langle R(\mathit{start}_{s_1}),\sigma'\rangle$}
	\end{prooftree}
	\begin{prooftree}
		\AxiomC{$ \llbracket c \rrbracket (\sigma) = \textit{false} $}
		\AxiomC{$\sigma' = \sigma$}
		\LeftLabel{$[\textit{ite}_{\textit{F}}]$}
		\BinaryInfC{$\langle R(\mathit{start}_{s}),\sigma\rangle \weirdarrow{}\langle R(\mathit{start}_{s'_1}),\sigma'\rangle$}
	\end{prooftree}

	Let $p_1$ be \while{s$_1;\ldots;$s$_k$} and let $\pv{s}$ be $\texttt{while}(Cond)\,\{p_1\}$. Instantiating the rules $while_T$ and $while_F$ with $p := \mathit{end}_s$ yields the two rules

	\begin{prooftree}
		\AxiomC{$ \llbracket c \rrbracket (\sigma) = \textit{true} $}
		\AxiomC{$\sigma' = \sigma$}
		\LeftLabel{[\textit{while}$_T$]}
		\BinaryInfC{$\langle R(tp_{s}(it^{s})),\sigma\rangle \weirdarrow{}\langle R(\mathit{start}_{s_1}),\sigma'\rangle$}
	\end{prooftree}
	\begin{prooftree}
		\AxiomC{$ \llbracket c \rrbracket (\sigma) = \textit{false} $}
		\AxiomC{$\sigma' = \sigma$}
		\LeftLabel{[\textit{while}$_F$]}
		\BinaryInfC{$\langle R(tp_{s}(it^{s})),\sigma\rangle \weirdarrow{} \langle R(\mathit{end}_{s}),\sigma'\rangle$}
	\end{prooftree}
	\caption{Small-step operational semantics using $\mathit{tp}$, $\mathit{start}$, $\mathit{end}$.}
	\label{fig:operational-semantics-timepoints}
\end{figure}

\subsection{$\whilelang$-Soundness}
Operational semantics describe the execution of a program. Such an execution correseponds to a model where the terms describing variable values are interpreted according to the states in the corresponding configurations. 

\begin{definition}[Execution-interpretation]
	Let $p$ be a program. For an arbitrary execution $E$ of $p$ containing configurations $\langle R(tp_1),\sigma_1 \rangle, \dots, \langle R(tp_n),\sigma_n \rangle$,
	let an \emph{execution-interpretation} be any interpretation $M$, such that for any integer/array variable $\pv{v}$ the term $v(tp_i)$ resp. $v(tp_i,pos)$ is interpreted as $\sigma_i(v)$ resp. $\sigma_i(v[pos])$ in $M$.
\end{definition}

With the above definition of execution-interpretations, soundness is captured as follows:

\begin{definition}[$\whilelang$-Soundness]
	Let $p$ be a program and let $A$ be a trace logic axiom.
	Then $A$ is called \emph{$\whilelang$-sound}, if for any execution-interpretation $M$ we have $M \vDash A$.
\end{definition}

We show that the axioms of trace logic are $\whilelang{}$-sound. % In order to simplify the presentation, we omit traces from the proof below. 

\begin{theorem}[$\whilelang$-Soundness of Semantics of $\whilelang$]
	For a given terminating program $p$, the axioms of Figure~\ref{fig:operational-semantics-timepoints} defining the semantics of $\whilelang$ are $\whilelang$-sound.
\end{theorem} 

\iffmcad
\begin{IEEEproof}
\else
\begin{proof}
\fi
Consider an arbitrary trace $tr$ and let $E$ be the execution denoted by $tr$. Let $M$ be an execution-interpretation of $E$. We proceed by structural induction on the program-structure with the induction hypothesis that for a subprogram $p' = s_1;\ldots;s_k$, if $\langle R(\mathit{start}_{s_1})\rangle$ is reachable our semantics $\llbracket p' \rrbracket$ instantiated with $tr$ are $\whilelang$-sound.

\balance

Case distinction on the type of the statement $s$:
\begin{itemize}
	\item Let $s$ be of the form $\skipOp$. Assume $\langle R(\mathit{start}_{s})\rangle$ is reachable. Then the only rule which applies is $skip$, so $\langle R(\mathit{end}_{s})\rangle$ is also reachable and $\sigma(R(\mathit{end}_{s})) = \sigma(R(\mathit{start}_{s}))$. By the definition of execution-interpretations, we therefore conclude that axiom~\ref{semantics_skip} is $\whilelang$-sound.

	\item Let $s$ be of the form $\mathit{v := e}$. Assume $\langle R(\mathit{start}_{s})\rangle$ is reachable. Then the only rule which applies is $asg$, so $\langle R(\mathit{end}_{s})\rangle$ is also reachable and $\sigma(R(\mathit{end}_{s})) = \sigma(R(\mathit{start}_{s}))[v \mapsto \llbracket e \rrbracket(\sigma(R(\mathit{start}_{s})))]$. By the definition of execution-interpretations, we therefore conclude that axiom~\ref{semantics_int_assign}  is $\whilelang$-sound.

	\item Let $s$ be of the form $\iteOp{c}{p_1}{p_2}$, where $p_1$ is $s_1;\ldots;s_k$. Assume $\langle R(\mathit{start}_{s})\rangle$ is reachable. Each axiom \ref{semantics_ite_1},\ref{semantics_ite_3} is an implication with $\llbracket c \rrbracket (\mathit{start}_{s})$ as positive premise, so assume that $\llbracket c \rrbracket (\mathit{start}_{s})$ holds in $M$.
	Then the only applicable rule is $ite_T$. 
	Therefore $\langle R(\mathit{start}_{s_1})\rangle$ is reachable and $\sigma(R(\mathit{start}_{s_1})) = \sigma(R(\mathit{start}_{s}))$. 
	From the latter fact and the definition of execution-interpretations we conclude that axiom \ref{semantics_ite_1} is $\whilelang$-sound. 
	Furthermore, since $p_1$ is a subprogram of $s$, we are able to combine the reachability of $\langle R(\mathit{start}_{p_1})\rangle$ with the induction hypothesis to derive that $\llbracket p_1\rrbracket$ is $\whilelang$-sound. In particular axiom \ref{semantics_ite_3} is $\whilelang$-sound.
	Analogously we are able to prove the $\whilelang$-soundness of axioms \ref{semantics_ite_2},\ref{semantics_ite_4}.

	\item Let $s$ be of the form $\whileOp{c}{p_1}$, where $p_1$ is $s_1;\ldots;s_k$. Assume $\langle R(\mathit{start}_{s})\rangle$ is reachable.
	Axiom \ref{semantics_while_1} and \ref{semantics_while_2} define $\mathit{lastIt}_s$ as the smallest iteration $it$ where $\llbracket c \rrbracket (tp_s(it))$ does not hold in $M$. Since we assume termination, such an iteration needs to exist, and in particular the definition is well-defined, so axiom \ref{semantics_while_1} and \ref{semantics_while_2} are $\whilelang$-sound.

	Now let $it$ be an arbitrary iteration such that $it<\mathit{lastIt}_s$ holds in $M$. By definition, for any iteration $it'<it$, we know that $\llbracket c \rrbracket (tp_s(it'))$ holds in $M$, in particular the only applicable rule for $\langle R(tp_s(it'))\rangle$ is $while_T$. Furthermore $\langle R(\mathit{start}_s)\rangle$ is reachable and equal to $\langle R(tp_s(0))\rangle$. Combining both facts we use a trivial sub-induction to conclude that $\langle R(tp_s(it))\rangle$ is reachable. Since $\llbracket c \rrbracket (tp_s(it))$ holds, the only applicable rule is $while_T$, so $\langle R(\mathit{start}_{s_1}(it))\rangle$ is reachable and $\sigma(R(\mathit{start}_{s_1}(it)))= \sigma(R(tp_{while}(it)))$. 
	From this we conclude that axiom \ref{semantics_while_3} is $\whilelang$-sound by the definition of execution-interpretations. Next, $p_1$ is a subexpression of $s$, so we are able to combine the reachability of $\langle R(\mathit{start}_{p_1}(it))\rangle$ with the induction hypothesis to conclude that $\llbracket p_1\rrbracket$ is $\whilelang$-sound. Therefore, axiom \ref{semantics_while_4} is $\whilelang$-sound.

	Finally, for any iteration $it<\mathit{lastIt}_s$, we know that $\llbracket c \rrbracket (tp_s(it))$ holds in $M$, so a similar trivial sub-induction yields that $\langle R(tp_s(\mathit{lastIt}_s))\rangle$ is reachable. By definition of $\mathit{lastIt}_s$, $\llbracket c \rrbracket (tp_s(it'))$ does not hold in $M$, so the only applicable rule is $while_F$. Therefore, $\langle R(\mathit{end}_s)\rangle$ is reachable and $\sigma(R(\mathit{end}_s)) = \sigma(R(tp_s(\mathit{lastIt}_{s})))$. In particular, Axiom \ref{semantics_while_5} is $\whilelang$-sound. 
\end{itemize}
Finally the initial state of an execution is reachable, so we can apply the induction hypothesis to establish the $\whilelang$-soundness of our semantics.
\iffmcad
\end{IEEEproof}
\else
\end{proof}
\fi

\fi

\end{document}